\documentstyle[aps,12pt,amssymb,epsfig]{revtex}

\begin{document}

\baselineskip=0.60cm
\newcommand{\ini}{\begin{equation}}
\newcommand{\fin}{\end{equation}}
\newcommand{\inir}{\begin{eqnarray}}
\newcommand{\finr}{\end{eqnarray}}
\newcommand{\inif}{\begin{figure}}
\newcommand{\finf}{\end{figure}}
\newcommand{\bc}{\begin{center}}
\newcommand{\ec}{\end{center}}

\def\ol{\overline}
\def\pa{\partial}
\def\ra{\rightarrow}
\def\ts{\times}
\def\df{\dotfill}
\def\bs{\backslash}
\def\dg{\dagger}

$~$

\hfill DSF-09/2002

\vspace{1 cm}

\centerline{{\bf SEESAW MECHANISM, BARYON ASYMMETRY}}
\centerline{{\bf AND NEUTRINOLESS DOUBLE BETA DECAY}}

\vspace{1 cm}

\centerline{\large{D. Falcone}}

\vspace{1 cm}

\centerline{Dipartimento di Scienze Fisiche, Universit\`a di Napoli,}
\centerline{Complesso di Monte S. Angelo, Via Cintia, Napoli, Italy}

\vspace{1 cm}

\begin{abstract}

\noindent
A simplified but very instructive analysis of the seesaw mechanism is here
performed. Assuming a nearly diagonal Dirac neutrino mass matrix,
we study the forms of the Majorana mass matrix of right-handed neutrinos,
which reproduce the effective mass matrix of left-handed neutrinos.
As a further step, the important effect of a non diagonal Dirac neutrino
mass matrix is explored.
The corresponding implications for the baryogenesis via leptogenesis and
for the neutrinoless double beta decay are reviewed.
We propose two distinct models where the baryon asymmetry is enhanced.

\end{abstract}

\newpage

\section{Introduction}

The seesaw mechanism \cite{ss} is a simple framework to account for the small
effective mass of the left-handed neutrino. It requires only a modest extension
of the minimal standard model, namely the addition of the right-handed
neutrino. As a consequence of this inclusion, both a Dirac mass term
for the neutrino and a Majorana mass term for the right-handed neutrino
are allowed. While the Dirac mass,
$m_{\nu}$, is expected to be of the same order of magnitude as the quark
or charged lepton mass, the Majorana mass of the right-handed neutrino,
$m_R$, is not constrained and may be very large. If this is the case,
a small effective Majorana mass for the left-handed neutrino,
$m_L \simeq (m_{\nu}/m_R)m_{\nu}$, is generated.

At the same time, the
out-of-equilibrium decay of the heavy neutrino can produce a baryon asymmetry
through the so-called baryogenesis via leptogenesis mechanism
\cite{fy,luty}. Hence, the
existence of the heavy Majorana neutrino may explain both the smallness
of the effective neutrino mass and the baryon asymmetry in the universe.
Moreover, the Majorana nature of the light neutrino, generated by the
Majorana nature of the heavy neutrino, allows the
neutrinoless double beta decay, because of lepton number violation at high
energy \cite{wein}. Thus the mass scale of the heavy neutrino could be a new
fundamental scale in physics.

For three generations of fermions, the light neutrino mass matrix $M_L$
as well as the baryon asymmetry $Y_B$ depend on the Dirac neutrino mass
matrix $M_{\nu}$ and the heavy neutrino mass matrix $M_R$. In this paper
we describe, in a simplified but instructive approach, the
structure of $M_{\nu}$ and $M_R$ within the seesaw mechanism and the
consequences for the baryon asymmetry generated in the baryogenesis via
leptogenesis mechanism and for the neutrinoless double beta decay.

As a first approximation, we assume a diagonal form for the Dirac neutrino
mass matrix, and then the effect of a non diagonal form is analyzed.
The prediction for the neutrinoless double beta decay can remain unchanged
because for a fixed $M_L$ and any choice of $M_{\nu}$ one can
find a certain $M_R$ which reproduces $M_L$ through the seesaw formula,
while the impact on the amount of baryon asymmetry is significant, because those
$M_R$ and $M_{\nu}$ determine $Y_B$ through the leptogenesis formula.

The outline of the paper is the following. In section II we give an approximate
description of the effective neutrino mass matrix. In sections III and IV we
briefly discuss the seesaw mechanism and the baryogenesis via leptogenesis,
respectively. In section V the
forms of the heavy neutrino mass matrix, and their implications for
the amount of baryon asymmetry
and the rate for neutrinoless double beta decay are reviewed,
according to different mass spectra of light neutrinos. In this section we
assume a diagonal Dirac neutrino mass matrix. The important effect of a
non diagonal Dirac neutrino mass matrix on the amount of baryon asymmetry
is explored in section VI, where we also propose two different models of
mass matrices, which produce an enhancement of the baryon asymmetry.
Finally, in section VII we summarize the subject.

\section{The effective neutrino mass matrix}

The lepton mixing matrix $U$ (called the Maki-Nakagawa-Sakata (MNS)
matrix \cite{mns}), which relates mass eigenstates to flavor eigenstates,
by means of the unitary transformation $\nu_{\alpha}=U_{\alpha i} \nu_i$
($\alpha=e,\mu,\tau$; $i=1,2,3$),
can be parametrized as the standard form of the Cabibbo-Kobayashi-Maskawa
(CKM) quark mixing matrix \cite{ckm}
(including a phase $\delta$) times a diagonal phase matrix
$P=\text{diag}(\text{e}^{\text{i}\varphi_1/2},\text{e}^{\text{i}\varphi_2/2},1)$.
The two phases $\varphi_1$ and $\varphi_2$ are present only if the effective
neutrino is a Majorana particle,
and thus they are sometimes called the Majorana phases, in contrast with
the phase $\delta$ which is called the Dirac phase.
Moreover, contrary to quark mixings, which are small, lepton mixings can be large.
In fact, the mixing of atmospheric neutrinos, related to $U_{\mu 3}$, is almost
maximal, while the mixing of solar neutrinos, related to $U_{e2}$, may be large
or small, although the large mixing is favoured \cite{bgp}.
In the case of double large mixing, the lepton mixing matrix is given by
\ini
U \simeq \left( \begin{array}{ccc}
\frac{1}{\sqrt2} & \frac{1}{\sqrt2} & \epsilon \\
-\frac{1}{2}(1+\epsilon) & \frac{1}{2}(1-\epsilon) & \frac{1}{\sqrt2} \\
\frac{1}{2}(1-\epsilon) & -\frac{1}{2}(1+\epsilon) & \frac{1}{\sqrt2}
\end{array} \right),
\fin
while for single large mixing it is given by
\ini
U \simeq \left( \begin{array}{ccc}
1 & 0 & \epsilon \\
-\frac{\epsilon}{\sqrt2} & \frac{1}{\sqrt2} & \frac{1}{\sqrt2} \\
-\frac{\epsilon}{\sqrt2} & -\frac{1}{\sqrt2} & \frac{1}{\sqrt2}
\end{array} \right),
\fin
see for example Ref.\cite{akh}.
As we said, the double large mixing is favoured. The mixing
$\epsilon=U_{e3}$ is very small,
$\epsilon \lesssim 0.1$, according to the result of the Chooz experiment
\cite{chooz}. If we call $D_L$ the diagonal matrix of light neutrino masses,
\ini
D_L = \text{diag} (m_1,m_2,m_3),
\fin
then, in the basis where the charged lepton mass matrix is diagonal,
$M_e=D_e$, we get
\ini
M_L = U D_L U^T.
\fin
In the flavor basis we have $M_L = U_e U D_L U^T U_e^T$,
where $U_e$ diagonalizes $M_e$ by $U_e^T M_e U_e$, where $M_e$ is here
supposed to be symmetric.
We do not consider general phases in $U$. However, we allow the masses
$m_1$ and $m_2$ to be both positive and negative, corresponding to phases
$\varphi_{1,2}=0$ and $\varphi_{1,2}=\pi$, respectively, in the lepton
mixing matrix. In such a way, we consider the two extreme cases for
$\varphi_{1,2}$ and the general case should be intermediate between them.

Let us call the elements of $M_L$ as $M_{\alpha \beta}$ with
$\alpha=e,\mu,\tau$ and $\beta=e,\mu,\tau$. Then, for the double large mixing
we have the approximate expressions
$$
M_{ee}=\frac{m_1}{2}+\frac{m_2}{2}+\epsilon^2 m_3
$$
$$
M_{e \mu}=-\frac{m_1}{2 \sqrt2}(1+\epsilon)+
\frac{m_2}{2 \sqrt2}(1-\epsilon)+\frac{\epsilon m_3}{\sqrt2}
$$
$$
M_{e \tau}=\frac{m_1}{2 \sqrt2}(1-\epsilon)-
\frac{m_2}{2 \sqrt2}(1+\epsilon)+\frac{\epsilon m_3}{\sqrt2}
$$
$$
M_{\mu \mu}=\frac{m_1}{4}(1+\epsilon)^2+
\frac{m_2}{4}(1-\epsilon)^2+\frac{m_3}{2}
$$
$$
M_{\mu \tau}=-\frac{m_1}{4}(1-\epsilon^2)-
\frac{m_2}{4}(1-\epsilon^2)+\frac{m_3}{2}
$$
$$
M_{\tau \tau}=\frac{m_1}{4}(1-\epsilon)^2+
\frac{m_2}{4}(1+\epsilon)^2+\frac{m_3}{2}
$$
and for the single large mixing the corresponding expressions
$$
M_{ee}=m_1 +\epsilon^2 m_3
$$
$$
M_{e \mu}=-\frac{\epsilon m_1}{\sqrt2} +\frac{\epsilon m_3}{\sqrt2}
$$
$$
M_{e \tau}=-\frac{\epsilon m_1}{\sqrt2} +\frac{\epsilon m_3}{\sqrt2}
$$
$$
M_{\mu \mu}=\frac{\epsilon^2 m_1}{2} +\frac{m_2}{2}+\frac{m_3}{2}
$$
$$
M_{\mu \tau}=\frac{\epsilon^2 m_1}{2} -\frac{m_2}{2}+\frac{m_3}{2}
$$
$$
M_{\tau \tau}=\frac{\epsilon^2 m_1}{2} +\frac{m_2}{2}+\frac{m_3}{2}.
$$
Of course, both $M_L$ and $M_R$ are symmetric matrices.
Note that for the single large mixing we have $M_{e \mu}=M_{e \tau}$ and
$M_{\mu \mu}=M_{\tau \tau}$.

The element $M_{ee}=U_{ei}^2 m_i$ is involved in the neutrinoless double beta decay.
The experimental upper bound for $|M_{ee}|$, obtained from non observation
of the process, is $|M_{ee}| < 0.38 h$ \cite{beta}, where the factor
$h=0.6-2.8$ \cite{fsv} is present because of the uncertainty in the calculation
of the nuclear matrix element. The recent positive evidence \cite{betasi}
is controversial \cite{fsv,betano} and is not used here. The parameter $M_{ee}$
is the unique element in $M_L$ which can be tested in a direct way, because
for the other elements the theoretical prediction is very much below the
experimental data \cite{rod}.

From the study of oscillations of atmospheric and solar neutrinos we know
that $|m_3^2-m_2^2| \gg |m_2^2-m_1^2|$, so that
there are three main mass spectra for the light neutrino, the normal
hierarchy $m_3^2 \gg m_2^2,m_1^2$, the inverse hierarchy
$m_1^2 \simeq m_2^2 \gg m_3^2$,
and the nearly degenerate spectrum $m_1^2 \simeq m_2^2 \simeq m_3^2$
(see Ref.\cite{af}).
In particular, from atmospheric oscillations we get $|m_3^2-m_2^2| \sim 10^{-3}$
eV$^2$, while from solar oscillations $|m_2^2-m_1^2| \sim 10^{-5}$ eV$^2$ for
the large mixing and $|m_2^2-m_1^2| \sim 10^{-6}$ eV$^2$ for the small mixing.
Then, with the normal hierarchy we get $m_3^2 \sim 10^{-3}$ eV$^2$, and
with the inverse hierarchy $m_{1,2}^2 \sim 10^{-3}$ eV$^2$.
For the nearly degenerate spectrum we expect $m_{1,2,3}^2$ around 1 eV$^2$.
In fact, the experimental upper bound on the parameter
$m_{\nu_e}=(U_{ei}^2 m_i^2)^{1/2}$, obtained from the endpoint energy
of electrons in single beta decay, is $m_{\nu_e} < 2.5$ eV \cite{beta1}.
In contrast with $M_{ee}$, in $m_{\nu_e}$ cancellations cannot occur.
If for the normal hierarchy also $m_2^2 \gg m_1^2$ is assumed, then
we get $m_2^2 \sim 10^{-5}$ eV$^2$ for the large mixing and
$m_2^2 \sim 10^{-6}$ eV$^2$ for the small mixing.

As a matter of fact, for solar neutrinos there are at least three
oscillation solutions with a large mixing angle \cite{bgp}:
the large mixing angle (LMA) matter oscillation with
$|m_2^2-m_1^2| \sim 10^{-5}$ eV$^2$, the low-mass (LOW) matter oscillation
with $|m_2^2-m_1^2| \sim 10^{-7}$ eV$^2$, and the vacuum oscillation (VO)
with $|m_2^2-m_1^2| \sim 10^{-10}$ eV$^2$. In our paper we refer mainly to the
LMA solution which is the most favoured \cite{bgp}.

Zeroth order forms for $M_L$ can be obtained by setting $\epsilon=0$
and $(m_1,m_2,m_3)$ equal to 
$(0,0,1)$, $(1,-1,0)$, $(1,1,0)$, $(1,1,1)$, $(-1,1,1)$, $(1,-1,1)$, 
$(-1,-1,1)$. We call these mass spectra $A$, $B_1$, $B_2$,
$C_0$, $C_1$, $C_2$, $C_3$, respectively. Of course, type $A$ is the normal
hierarchy, type $B$ the inverse hierarchy, and type $C$ nearly degenerate. 
In the present paper we write only the zeroth order form of $M_L$, although
we study the full element $M_{ee}$. For a more detailed description of $M_L$
see for example \cite{akh}.

\section{Seesaw mechanism}

Since in the standard model with right-handed neutrinos the Dirac neutrino
mass is generated on the same footing as the up quark masses,
and the charged lepton masses on the same footing as the down quark masses,
for the corresponding mass matrices we first assume the
relations $M_{\nu} \simeq M_u$ and $M_{e} \simeq M_d$, that is 
\ini
M_{\nu} \simeq \text{diag} (m_u,m_c,m_t),
\fin
and $M_e \simeq \text{diag} (m_e,m_{\mu},m_{\tau})$,
where we neglect the small Dirac lepton mixing analogous to the quark mixing.
In other terms, we  set $U_e \simeq \openone$ and $U_{\nu} \simeq \openone$,
where $U_e$ diagonalizes $M_e$ and $U_{\nu}$ diagonalizes $M_{\nu}$.
As a matter of fact, the mass hierarchy of charged leptons is similar to that
of down quarks. See Ref.\cite{ff} for a summary on quark and lepton mass
matrices. Because of the seesaw mechanism, we have
\ini
M_L \simeq M_{\nu} M_R^{-1} M_{\nu},
\fin
where the heavy neutrino masses $M_1$, $M_2$, $M_3$
have to be much larger than the elements of the matrix $M_{\nu}$, which is here
supposed to be symmetric.
Inverting such a formula, the heavy neutrino mass matrix can be achieved,
\ini
M_R \simeq M_{\nu} M_L^{-1} M_{\nu}.
\fin
Since $M_L^{-1}=U D_L^{-1} U^T$, we can get $M_L^{-1}$ from $M_L$ by
changing $m_i$ with $1/m_i$ in (4).
We must keep in mind that for the inverse hierarchy
the seesaw mechanism implies a cancellation of the Dirac hierarchy for the
third and second
generation, and for the nearly degenerate spectrum also with the first
generation, which seems unnatural, especially for the VO solution.
Note that for zero mixing we have $m_1=m_u^2/M_1$,
$m_2=m_c^2/M_2$ and $m_3=m_t^2/M_3$. 

In this paper we follow a kind of inverse or bottom-up approach,
namely we will determine $M_R$ from $M_L$ (and $M_{\nu}$).
In the alternative direct or top-down approach both $M_R$ and $M_{\nu}$ are
obtained from a theoretical framework and the inferred $M_L$ is matched to
neutrino phenomenology. We would like to stress that for any precise
choice of $M_{\nu}$ we can reproduce any form of $M_L$ by adjusting $M_R$.
However, as we will see in the following sections, the form of both
$M_{\nu}$ and $M_R$ has a crucial impact on the amount of baryon asymmetry
generated in the leptogenesis mechanism, so that we can use the constraint
from baryogenesis to study $M_R$ and $M_{\nu}$. Unfortunately, we are not yet
able to determine the mass matrices with precision. Nevertheless, some important
considerations, involving also the neutrinoless double beta decay,
can be done and this would be a central issue of our paper.

\section{Baryogenesis from leptogenesis}

A baryon asymmetry can be generated from a lepton asymmetry \cite{fy}.
In fact, this lepton asymmetry is produced by the out-of-equilibrium CP-violating
decay of heavy neutrinos. The electroweak sphalerons \cite{krs},
which violates $B+L$
but conserve $B-L$, transform part of this asymmetry into a baryon asymmetry.
Then, the baryon asymmetry, defined as
$Y_B=(n_B-n_{\ol{B}})/7n_{\gamma}=\eta/7$, where $n_B$, $n_{\ol{B}}$,
$n_{\gamma}$ are number densities
and $\eta$ is the baryon-to-photon ratio,
can be written as (see Ref.\cite{ft,crv} and references therein)
\ini
Y_B \simeq \frac{1}{2}~\frac{1}{g^*}~d~\epsilon_1,
\fin
with the CP violating asymmetry in the decay of the lightest heavy neutrino
with mass $M_1 \ll M_2 < M_3$ given by
\ini
\epsilon_1 \simeq \frac{3}{16 \pi v^2}
\left[\frac{[(M_D^{\dg} M_D)_{12}]^2}{(M_D^{\dg} M_D)_{11}}
\frac{M_1}{M_2}+\frac{[(M_D^{\dg} M_D)_{13}]^2}{(M_D^{\dg}
M_D)_{11}}
\frac{M_1}{M_3} \right],
\fin
where $M_D=M_{\nu} U_R$ and $U_R^T M_R U_R=D_R$,
and $v \simeq m_t$ is the VEV of the Higgs doublet.
The lightest heavy neutrino is in equilibrium during the decays of the two
heavier ones, washing out the lepton asymmetry generated by them.
The factor 1/2 represents the
part of the lepton asymmetry converted into a baryon asymmetry \cite{htks}.
The parameter $g^* \simeq 100$
is the number of light degrees of freedom in the theory.
Finally, the quantity $d$ is a dilution factor which mostly depends on
the mass parameter
\ini
\tilde{m}_1 = \frac{(M_D^{\dg} M_D)_{11}}{M_1},
\fin
although for high values of $\tilde{m}_1$ some dependence on $M_1$ shows up. 
Minor dilution, $d$ of order $10^{-1}$, is obtained for
$10^{-5} \text{eV} \lesssim \tilde{m}_1 \lesssim 10^{-2} \text{eV}$,
while outside this range the dilution grows (that is $d$ diminishes)
\cite{bp,hk}.
In fact, if $\tilde{m}_1$ is too small, it is not possible to produce a
sufficient number of heavy neutrinos at high temperature, while if $\tilde{m}_1$
is too large, the washout effect of lepton number violating scatterings is too
strong and destroys the generated asymmetry.
In order to be consistent with primordial nucleosynthesis, the baryon asymmetry
$Y_B$ must be in the range $10^{-11}-10^{-10}$ \cite{olive}.
At best, $Y_B$ is smaller than $\epsilon_1$ by three orders of magnitude.
It is clear that when we obtain $M_R$ from $M_{\nu}$ and $M_L$ through the
inverse seesaw formula (7), a determination of the baryon asymmetry is also
achieved. In the following two sections we try a partial selection of
mass matrices using the bound on the amount of baryon asymmetry.

\section{Symplified analysis of the seesaw mechanism}

In this section we determine the structure of the heavy neutrino mass matrix
by assuming the diagonal form (5) for the Dirac neutrino mass matrix.
In the next section the effect of a non diagonal form is discussed.
In any subsection we write the zeroth order form of $M_L$, according to the
three possible hierarchies of light neutrino masses, and then we study the
heavy neutrino mass matrix and the implications for the baryon asymmetry and
for the neutrinoless double beta decay.

\subsection{Normal hierarchy}

For both kinds of mixings the zeroth order form for $M_L$ is given by
\ini
M_L \sim \left( \begin{array}{ccc}
0 & 0 & 0 \\ 0 & \frac{1}{2} & \frac{1}{2} \\ 0 & \frac{1}{2} & \frac{1}{2}
\end {array} \right) m_3,
\fin
where a dominant block in the $\mu-\tau$ sector appears
(see for example \cite{by}). The overall scale is $m_3 \sim 10^{-2}-10^{-1}$ eV.

\subsubsection{Double large mixing}

If there is full hierarchy of light neutrino masses, $m_3 \gg m_2 \gg m_1$,
the matrix $M_L^{-1}$ takes a nearly democratic form \cite{fal}, so that
\ini
M_R \simeq \left( \begin{array}{ccc}
m_u^2 & -m_u m_c & m_u m_t \\
-m_u m_c & m_c^2 & -m_c m_t \\
m_u m_t & -m_c m_t & m_t^2
\end {array} \right) \frac{1}{m_1}.
\fin
This matrix is diagonalized by the rotation
\ini
U_R \simeq \left( \begin{array}{ccc}
1 & -\frac{m_u}{m_c} & \frac{m_u}{m_t} \\
\frac{m_u}{m_c} & 1 & -\frac{m_c}{m_t} \\
\frac{m_u}{m_t} & \frac{m_c}{m_t} & 1
\end {array} \right),
\fin
with eigenvalues $M_1 \simeq m_u^2/m_1$, $M_2 \simeq m_c^2/m_1$,
$M_3 \simeq m_t^2/m_1$. For the full hierarchy and large solar mixing
we have $m_2^2 \sim 10^{-5}$ eV$^2$, hence $m_1 \lesssim 10^{-3}$ eV
and the overall scale is $M_R \sim m_t^2/m_1 \gtrsim 10^{16}$ GeV. 
In this section we take the full hierarchy as reference model.

If $m_3 \gg m_2 \simeq m_1$ (partially degenerate spectrum), we get
\ini
M_R \simeq \left( \begin{array}{ccc}
m_u^2 & -\epsilon m_u m_c & -\epsilon m_u m_t \\
-\epsilon m_u m_c & m_c^2 & -m_c m_t \\
-\epsilon m_u m_t & -m_c m_t & m_t^2
\end {array} \right) \frac{1}{m_1},
\fin
where very small elements $M_{R12}$ and $M_{R13}$ appear. The overall scale
and the heavy masses do not change with respect to the full hierarchical case.

The further condition $m_2 < 0$ leads to the form
\ini
M_R \simeq \left( \begin{array}{ccc}
\epsilon^2 \frac{m_1}{m_3} m_u^2 & -m_u m_c & m_u m_t \\
-m_u m_c & (\frac{m_1}{m_3}+\epsilon) m_c^2 & \frac{m_1}{m_3} m_c m_t \\
m_u m_t & \frac{m_1}{m_3} m_c m_t & (\frac{m_1}{m_3}-\epsilon) m_t^2
\end {array} \right) \frac{1}{m_1}.
\fin
If $m_1 \gg \epsilon m_3$ the overall scale is lowered to
$M_R \sim m_t^2/m_3 \sim 10 ^{14}$ GeV.
For $m_1 \simeq \epsilon m_3$ we get the interesting form
\ini
M_R \simeq \left( \begin{array}{ccc}
\epsilon^3 m_u^2 & -m_u m_c & m_u m_t \\
-m_u m_c & 2 \epsilon m_c^2 & \epsilon m_c m_t \\
m_u m_t & \epsilon m_c m_t & 0
\end {array} \right) \frac{1}{m_1}.
\fin
We stress the sharp difference between matrix (16) and matrices (12) or (14).
While in (12) and (14) the largest element is $M_{R33}$,
in (16) it is $M_{R13}$ or $M_{R23}$. In the first case the structure of $M_R$
is roughly similar to the Dirac neutrino mass matrix $M_{\nu}$ in (5), that is
a nearly diagonal form. In the second case, $M_R$ is roughly offdiagonal. As a
consequence, also the overall mass scale is different,
$m_t^2/m_1 \gtrsim 10^{16}$ GeV for the nearly diagonal form and
$m_u m_t/m_1 \gtrsim 10^{11}$ GeV for the nearly offdiagonal form.

Let us discuss the implications for the baryogenesis via leptogenesis
and for the neutrinoless double beta decay.
The baryon asymmetry for the full normal hierarchy is $Y_B \sim 10^{-16}$.
Due to the suppression of $M_{R12}$ and $M_{R13}$, the baryon asymmetry is
smaller in the partially degenerate spectrum than in the full normal hierarchy.
In fact, the relation $M_D^{\dg} M_D \simeq M_R m_1$ holds in both cases,
since $M_D^{\dg} M_D$ and $M_R m_1$ are diagonalized by the same $U_R$ with
the same eigenvalues.
For the nearly offdiagonal form, the baryon asymmetry is enhanced to a
sufficient level, due to the moderate hierarchy within $M_R$.
In the neutrinoless double beta decay we get
$M_{ee} \simeq m_2 \sim 10^{-3}-10^{-2}$ eV if $m_2 >0$, and
$M_{ee} \simeq (m_2^2-m_1^2)/m_{1,2} \sim 10^{-4}-10^{-3}$ eV if $m_2 < 0$.
Therefore, the nearly offdiagonal form for $M_R$ tends to enhance $Y_B$ but to
suppress $M_{ee}$. Thus we have found that if the diagonal $M_{\nu}$ in (5)
is assumed then in order to get sufficient baryon asymmetry the matrix $M_R$
must be roughly offdiagonal, which leads to a negative $m_2$ and an approximate
prediction for $M_{ee}$, smaller by one order with respect to the case of a
roughly diagonal $M_R$.
The negative value for $m_2$ could be an indication
that phases $\varphi_1$ and $\varphi_2$ are very different from each other. 

\subsubsection{Single large mixing}

In this case we have the form
\ini
M_R \simeq \left( \begin{array}{ccc}
m_u^2 & -\epsilon m_u m_c & -\epsilon m_u m_t \\
-\epsilon m_u m_c & (\frac{m_1}{m_2}+\epsilon^2) m_c^2 & (\frac{m_1}{m_2}-\epsilon^2) m_c m_t \\
-\epsilon m_u m_t & (\frac{m_1}{m_2}-\epsilon^2) m_c m_t & (\frac{m_1}{m_2}+\epsilon^2) m_t^2
\end {array} \right) \frac{1}{m_1},
\fin
and for $m_1 \gg \epsilon^2 m_2$
the structure of $M_R$ is nearly diagonal with the overall scale given by
$m_t^2/m_2 \sim 10^{15}$ GeV. The baryon asymmetry can get a moderate
enhancement, because of the factor $m_2/m_1$, and
$M_{ee} \simeq m_1 \lesssim 10^{-4}$ eV.
If $m_1 \simeq \epsilon^2 m_2$ the element $M_{R23}$ vanishes.
For $m_2 <0$ and $m_1 \simeq -\epsilon^2 m_2$, we have a vanishing $M_{R33}$,
leading to a nearly offdiagonal form at a lower scale.

\subsection{Inverse hierarchy}

The zeroth order form of $M_L$ for spectrum $B_1$ is given by
\ini
M_L \sim \left( \begin{array}{ccc}
0 & -\frac{1}{\sqrt2} & \frac{1}{\sqrt2} \\
-\frac{1}{\sqrt2} & 0 & 0 \\ \frac{1}{\sqrt2} & 0 & 0
\end {array} \right) m_1,
\fin
\ini
M_L \sim \left( \begin{array}{ccc}
1 & 0 & 0 \\ 0 & -\frac{1}{2} & \frac{1}{2} \\ 0 & \frac{1}{2} & -\frac{1}{2}
\end {array} \right) m_1,
\fin
according to the double or single large mixing. For spectrum $B_2$ we have
\ini
M_L \sim \left( \begin{array}{ccc}
1 & 0 & 0 \\ 0 & \frac{1}{2} & -\frac{1}{2} \\ 0 & -\frac{1}{2} & \frac{1}{2}
\end {array} \right) m_1
\fin
for both cases. The overall scale is $m_{1,2} \sim 10^{-2}-10^{-1}$ eV.
This is the same as the normal hierarchy, because both are determined by
atmospheric oscillations.

\subsubsection{Double large mixing}

The heavy neutrino mass matrix for spectrum $B_2$ is given by
\ini
M_R \simeq \left( \begin{array}{ccc}
\frac{m_3}{m_1} m_u^2 & \epsilon m_u m_c & \epsilon m_u m_t \\
\epsilon m_u m_c & m_c^2 & m_c m_t \\
\epsilon m_u m_t & m_c m_t & m_t^2
\end {array} \right) \frac{1}{m_3}.
\fin

If $m_2 <0$, spectrum $B_1$, we get
\ini
M_R \simeq \left( \begin{array}{ccc}
\epsilon^2 m_u^2 & -\frac{m_3}{m_1} m_u m_c & \frac{m_3}{m_1} m_u m_t \\
-\frac{m_3}{m_1} m_u m_c & m_c^2 & m_c m_t \\
\frac{m_3}{m_1} m_u m_t & m_c m_t & m_t^2
\end {array} \right) \frac{1}{m_3}.
\fin
In the inverse pattern there is  stronger hierarchy in $M_R$ with
respect to the full normal pattern, see the first row and column
in (21) and (22).
The form of $M_R$ is always nearly diagonal and the offdiagonal form cannot
be realized. The mass scale is $M_R \sim m_t^2/m_3 \gtrsim 10^{15}$ GeV.
Note also the difference between
the inverse hierarchy and the partially degenerate spectrum in the
element $M_{R11}$, which is responsible for the inversion of the light neutrino
masses $m_{1,2}$ and $m_3$.

As a consequence, the baryon asymmetry is even smaller
than in the partially degenerate spectrum with $m_2 >0$,
because also $M_1$ is suppressed.
Instead, the rate for neutrinoless double beta decay, related to $M_{ee}$,
can be enhanced. In particular, for spectrum $B_2$ we have
$M_{ee} \simeq m_{1,2} \sim 10^{-2}-10^{-1}$ eV, which is by one order higher
than for the full normal hierarchy.
For spectrum $B_1$
we obtain $M_{ee} \simeq (m_2^2-m_1^2)/m_{1,2} \sim 10^{-4}-10^{-3}$ eV.

\subsubsection{Single large mixing}

In this case the heavy neutrino mass matrix is the same as for the previous case
with $m_2 >0$, and $M_{ee} \simeq m_1 \sim 10^{-2}-10^{-1}$ eV.

\subsection{Nearly degenerate spectrum}

Here the light masses are around 1 eV.
The zeroth order form of $M_L$ for spectrum $C_0$ is diagonal,
\ini
M_L \sim \left( \begin{array}{ccc}
1 & 0 & 0 \\ 0 & 1 & 0 \\ 0 & 0 & 1
\end {array} \right) m_1,
\fin
for both kinds of mixing. For spectrum $C_1$ we have
\ini
M_L \sim \left( \begin{array}{ccc}
0 & \frac{1}{\sqrt2} & -\frac{1}{\sqrt2} \\
\frac{1}{\sqrt2} & \frac{1}{2} & \frac{1}{2} \\
-\frac{1}{\sqrt2} & \frac{1}{2} & \frac{1}{2}
\end {array} \right) m_1,
\fin
\ini
M_L \sim \left( \begin{array}{ccc}
-1 & 0 & 0 \\ 0 & 1 & 0 \\ 0 & 0 & 1
\end {array} \right) m_1,
\fin
and for spectrum $C_2$
\ini
M_L \sim \left( \begin{array}{ccc}
0 & -\frac{1}{\sqrt2} & \frac{1}{\sqrt2} \\
-\frac{1}{\sqrt2} & \frac{1}{2} & \frac{1}{2} \\
\frac{1}{\sqrt2} & \frac{1}{2} & \frac{1}{2}
\end {array} \right) m_1,
\fin
\ini
M_L \sim \left( \begin{array}{ccc}
1 & 0 & 0 \\ 0 & 0 & 1 \\ 0 & 1 & 0
\end {array} \right) m_1,
\fin
according to the double or single large mixing.
Finally, for spectrum $C_3$
\ini
M_L \sim \left( \begin{array}{ccc}
-1 & 0 & 0 \\ 0 & 0 & 1 \\ 0 & 1 & 0
\end {array} \right) m_1
\fin
for both kinds of mixing. The overall scale is $m_{1,2,3} \sim 0.1-1$ eV,
determined by single beta decay.
In the nearly degenerate spectrum several delicate
cancellations among the terms of $M_{ee}$
may occur and our study of the heavy neutrino mass matrix is only
indicative.

\subsubsection{Double large mixing}

The heavy neutrino mass matrix for the spectra $C_0$, $C_1$, $C_2$, $C_3$ is
respectively given by the forms
\ini
M_R \simeq \left( \begin{array}{ccc}
m_u^2 & \rho m_u m_c & \rho m_u m_t \\
\rho m_u m_c & m_c^2 & \epsilon^2 m_c m_t \\
\rho m_u m_t & \epsilon^2 m_c m_t & m_t^2
\end {array} \right) \frac{1}{m_1},
\fin
\ini
M_R \simeq \left( \begin{array}{ccc}
\epsilon^2 m_u^2 & m_u m_c & -m_u m_t \\
m_u m_c & m_c^2 & m_c m_t \\
-m_u m_t & m_c m_t & m_t^2
\end {array} \right) \frac{1}{m_1},
\fin
\ini
M_R \simeq \left( \begin{array}{ccc}
\epsilon^2 m_u^2 & -m_u m_c & m_u m_t \\
-m_u m_c & m_c^2 & m_c m_t \\
m_u m_t & m_c m_t & m_t^2
\end {array} \right) \frac{1}{m_1},
\fin
\ini
M_R \simeq \left( \begin{array}{ccc}
-m_u^2 & \epsilon m_u m_c & \epsilon m_u m_t \\
\epsilon m_u m_c & -\epsilon^2 m_c^2 & m_c m_t \\
\epsilon m_u m_t & m_c m_t & -\epsilon^2 m_t^2
\end {array} \right) \frac{1}{m_1}.
\fin
In (29), and (34) below, we write $\rho$ whenever a cancellation at the
level of $\epsilon$ occurs. We see that matrices (29), (30), (31) are roughly
close to the diagonal form, while (32) is not. The overall mass scale is
$M_R \sim 10^{13}$ GeV.

For spectra $C_0$ and $C_3$
the baryon asymmetry is generally suppressed or much suppressed with respect to
the full normal hierarchy, while for spectra $C_1$ and $C_2$ it is comparable.
Instead, the rate for neutrinoless double beta decay can be further enhanced
with respect to the inverse hierarchy. In particular, for spectra
$C_0$ and $C_3$ we have $M_{ee} \simeq m_1 \sim 0.1-1$ eV, by one order higher
than for the spectrum $B_2$.
For spectra $C_1$ and $C_2$ we have cancellations leading to
$M_{ee} \sim 10^{-5}-10^{-4}$ eV.

\subsubsection{Single large mixing}

For spectra $C_0$ and $C_3$ we have the same matrices as for the previous
mixing. For spectra $C_1$ and $C_2$ we get
\ini
M_R \simeq \left( \begin{array}{ccc}
-m_u^2 & \epsilon m_u m_c & \epsilon m_u m_t \\
\epsilon m_u m_c & m_c^2 & -\epsilon^2 m_c m_t \\
\epsilon m_u m_t & -\epsilon^2 m_c m_t & m_t^2
\end {array} \right) \frac{1}{m_1},
\fin
\ini
M_R \simeq \left( \begin{array}{ccc}
m_u^2 & \rho m_u m_c & \rho m_u m_t \\
\rho m_u m_c & \epsilon^2 m_c^2 & m_c m_t \\
\rho m_u m_t & m_c m_t & \epsilon^2 m_t^2
\end {array} \right) \frac{1}{m_1}.
\fin
In this two cases the baryon asymmetry is suppressed or much suppressed
with respect to the full normal hierarchy, and $M_{ee} \sim 0.1-1$ eV.

\section{Seesaw mechanism and baryon asymmetry}

In this section, instead of the diagonal form, we take the realistic
mass matrices, expressed in terms of the Cabibbo parameter $\lambda=0.22$
and the overall mass scale,
\ini
M_{e} \sim \left( \begin{array}{ccc}
\lambda^{6} & \lambda^3 & \lambda^{5} \\
\lambda^3 & \lambda^2 & \lambda^2 \\
\lambda^{5} & \lambda^2 & 1
\end{array} \right)m_{b},
\fin

\ini
M_{\nu} \sim \left( \begin{array}{ccc}
\lambda^{12} & \lambda^6 & \lambda^{10} \\
\lambda^6 & \lambda^4 & \lambda^4 \\
\lambda^{10} & \lambda^4 & 1
\end{array} \right)m_t.
\fin
These forms can be motivated
by an $U(2)$ horizontal symmetry, see Ref.\cite{falco} and references therein.
Again we neglect $U_e$ with respect to $U$.
However, the effect of $U_{\nu} \ne \openone$ is crucial.
Of course, $M_L^{-1}$ can be obtained from the previous section by deleting
the quark masses in $M_R$.
In the following calculation
we will assume that no cancellations occur between two quantities of the same
order in $\lambda$.

For the full normal hierarchy and double large mixing we get
\ini
M_R \sim \left( \begin{array}{ccc}
\lambda^{12} & \lambda^{10} & \lambda^6 \\
\lambda^{10} & \lambda^8 & \lambda^4 \\
\lambda^6 & \lambda^4 & 1
\end{array} \right) \frac{m_t^2}{m_1},
\fin
diagonalized by
\ini
U_R \sim \left( \begin{array}{ccc}
1 & \lambda^{2} & \lambda^6 \\
-\lambda^{2} & 1 & \lambda^4 \\
\lambda^6 & -\lambda^4 & 1
\end{array} \right),
\fin
with eigenvalues $M_3 \sim {m_t^2}/{m_1}$, $M_2 \sim \lambda^8 M_3$,
$M_1 \sim \lambda^{12} M_3$, and consistent with the $U(2)$ horizontal symmetry
\cite{falco}.
The baryon asymmetry is enhanced with respect to the diagonal case but
remains too small, $Y_B \sim 10^{-14}$.
In fact, the relation $M_D^{\dg} M_D \sim M_R m_1$ holds, and one
obtains
\ini
\epsilon_1 \sim \frac{3}{16 \pi}
\left( \frac{\lambda^{20}}{\lambda^{12}}~\lambda^4+
\frac{\lambda^{12}}{\lambda^{12}}~\lambda^{12} \right) \sim 10^{-10},
\fin
and $\tilde{m}_1 \sim m_1$.
The same $M_R$ and $Y_B$ come out for the partially degenerate spectrum
with $m_2 >0$, the inverse hierarchy, and the nearly degenerate spectra
$C_1$ and $C_2$, although the scale of $M_R$ is changed accordingly.

For the partially degenerate spectrum with $m_2 <0$, assuming for example
both $\epsilon \sim \lambda^4$ and $M_{R33} \sim \lambda^{12} {m_t^2}/{m_1}$,
we obtain
\ini
M_R \sim \left( \begin{array}{ccc}
\lambda^{16} & \lambda^{12} & \lambda^{10} \\
\lambda^{12} & \lambda^{10} & \lambda^6 \\
\lambda^{10} & \lambda^6 & \lambda^8
\end{array} \right) \frac{m_t^2}{m_1}.
\fin
By considering $M_R^{\dg} M_R$, one finds that $M_R$ is diagonalized by
\ini
U_R \sim \left( \begin{array}{ccc}
1 & \lambda^{4} & \lambda^6 \\
-\lambda^{4} & 1 & \lambda^2 \\
\lambda^6 & -\lambda^2 & 1
\end{array} \right),
\fin
with eigenvalues $M_3 \sim \lambda^6 {m_t^2}/{m_1}$, $M_2 \sim M_3$,
$M_1 \sim \lambda^{4} M_3$.
Then we obtain
\ini
\epsilon_1 \sim \frac{3}{16 \pi}
\left( \frac{\lambda^{16}}{\lambda^{12}}~\lambda^4+
\frac{\lambda^{12}}{\lambda^{12}}~\lambda^{4} \right) \sim 10^{-4},
\fin
and $\tilde{m}_1 \sim \lambda^2 m_1$, so that a sufficient amount
of baryon asymmetry can be easily achieved.
Note that here the dominant term in the leptogenesis formula is the second one.
The mass scale in (37) is given by $m_t^2/m_1 \gtrsim 10^{16}$ GeV, and in (40)
by $\lambda^6 m_t^2/m_1 \gtrsim 10^{12}$ GeV.

For spectrum $C_0$ we have the nearly diagonal form
\ini
M_R \sim \left( \begin{array}{ccc}
\lambda^{12} & \lambda^{10} & \lambda^{10} \\
\lambda^{10} & \lambda^8 & \lambda^4 \\
\lambda^{10} & \lambda^4 & 1
\end{array} \right) \frac{m_t^2}{m_1},
\fin
and for spectrum $C_3$
\ini
M_R \sim \left( \begin{array}{ccc}
\lambda^{16} & \lambda^{10} & \lambda^6 \\
\lambda^{10} & \lambda^8 & \lambda^4 \\
\lambda^6 & \lambda^4 & \lambda^4
\end{array} \right) \frac{m_t^2}{m_1},
\fin
The baryon asymmetry is very small in the case $C_0$ and moderate in the 
case $C_3$.

For the single large mixing the results are roughly similar to the double large
mixing. Few changes can be easily shown and are not discussed here.
Thus we see that using non diagonal Dirac neutrino mass matrices
and doing an order-of-magnitude analysis lead to few forms for the heavy
neutrino mass matrix. Then the two questions of determining the symmetry
generating mass matrices and discovering their coefficient should be addressed.
For example, horizontal $U(1)$ or $U(2)$ symmetries can be used.
This is a fundamental subject that we will try to discuss elsewhere.
We need not start from matrices (35), (36) but other forms are possible
as well \cite{dv}.

A simple way to enhance the baryon asymmetry is by means of a quite
moderate hierarchy in the Dirac neutrino mass matrix, that is for its
eigenvalues and/or for its mixing angles \cite{falc,bs}. This is quite evident
from the leptogenesis formula. For example, instead of (36), one can adopt
\ini
M_{\nu} \sim \left( \begin{array}{ccc}
\lambda^{6} & \lambda^3 & \lambda^{5} \\
\lambda^3 & \lambda^2 & \lambda^2 \\
\lambda^{5} & \lambda^2 & 1
\end{array} \right)m_{t},
\fin
that is a matrix similar to charged lepton masses but with the same overall scale
as up quark masses. Let us discuss this issue. In section III we have assumed
$M_e \sim M_d$ and $M_{\nu} \sim M_u$. This is the simplest hypothesis
within the standard model, and in the supersymmetric model can be motivated
by the fact that the two pairs $M_{e,d}$ and $M_{\nu,u}$ are generated by
two distinct Higgs doublets.
However, Yukawa couplings for the Dirac neutrino can be very
different from the Yukawa couplings for the up quarks. If this case occurs,
the hierarchy of masses and mixings in the Dirac neutrino sector
can be very different from the mass hierarchy of up quarks and the CKM quark
mixing, respectively. When we take matrices (35) and (45) we obtain
\ini
M_R \sim \left( \begin{array}{ccc}
\lambda^{6} & \lambda^{5} & \lambda^3 \\
\lambda^{5} & \lambda^4 & \lambda^2 \\
\lambda^3 & \lambda^2 & 1
\end{array} \right) \frac{m_t^2}{m_1},
\fin
\ini
\epsilon_1 \sim \frac{3}{16 \pi}
\left( \frac{\lambda^{10}}{\lambda^{6}}~\lambda^2+
\frac{\lambda^{6}}{\lambda^{6}}~\lambda^{6} \right) \sim 10^{-6},
\fin
and a sufficient amount of baryon asymmetry, $Y_B \sim 10^{-10}$.
Both the internal hierarchy and the
overall scale of $M_{\nu}$ are important. In fact, if the overall scale in (45)
is $m_{b,\tau}$ instead of $m_t$, the baryon asymmetry again suppressed \cite{falc}.
The relation $M_{\nu} \sim (\tan \beta) M_e$ can be obtained within supersymmetric
left-right models \cite{bdm}. For very large $\tan \beta$ we get a Dirac
neutrino mass matrix similar to (45).

\section{Discussion}

Assuming a nearly diagonal mass matrix for Dirac neutrinos, we have studied
the structure of the Majorana mass matrix of right-handed neutrinos within
the seesaw framework, and its implications for the baryogenesis via leptogenesis
and for the neutrinoless double beta decay. Then we have explored the effect
of a non diagonal mass matrix for Dirac neutrinos.
In this context, we find few possibilities to obtain a sufficient level of
baryon asymmetry. The only case when the asymmetry is large should be the
nearly offdiagonal form of $M_R$.
Usually, the behaviour of $M_{ee}$ is opposite to $Y_B$, namely when
$Y_B$ is suppressed $M_{ee}$ is enhanced and viceversa. For the offdiagonal
form we get $M_{ee} \sim 10^{-4}-10^{-3}$ eV.
In the supersymmetric formula for the leptogenesis,
the baryon asymmetry is only slightly enhanced
\cite{ft,bp}. In fact, although there are new decay channels, also the washout
process is stronger.

When a moderate hierarchy in $M_{\nu}$ is adopted, as in (45),
then $M_R$ need not be close to the offdiagonal form. Therefore, instead of
$M_{ee} \sim 10^{-4}-10^{-3}$ eV, we can yield the value
$M_{ee} \sim 10^{-3}-10^{-2}$ eV for the normal hierarchy and
$M_{ee} \sim 10^{-2}-10^{-1}$ eV for the inverse hierarchy.
Note that these predictions cover three different ranges of values for $M_{ee}$,
so that informations from the neutrinoless double beta decay could clarify
the structure of fermion mass matrices, if the leptogenesis mechanism is valid.

As a conclusion, we find that, if $Y_B$ has to be within the allowed range,
then retaining quark-lepton mass relations $M_e \sim M_d$ and
$M_{\nu} \sim M_u$ leads to the roughly offdiagonal form for $M_R$ and the
prediction $M_{ee} \sim 10^{-4}-10^{-3}$ eV. If $M_{\nu} \sim M_u$ is not
true, then $M_R$ can be roughly close to the diagonal form and $M_{ee}$
larger than $10^{-3}$ eV, with $Y_B$ in the allowed range.
This is the central result of our paper. We have proposed such two different
kinds of model in the previous section.

It has been suggested \cite{fal} that the nearly offdiagonal form for $M_R$ is
consistent with the nonsupersymmetric unified model $SO(10)$
with an intermediate symmetry breaking scale, where the heavy neutrino mass is
generated, while the nearly diagonal form for $M_R$ is consistent with
the supersymmetric version without such an intermediate scale,
the heavy neutrino mass being generated at the unification scale.
In both cases, quark-lepton symmetry is valid. Then, we ask if the framework
described here could be embedded in unified models.
On the one hand, the leptogenesis scenario could work
within unified models \cite{pil}.
On the other hand, unified model have not yet received decisive support from
the experimental detection of proton decay, so that we find attractive
to keep the minimal scenario, as was the motivation of the original paper
on the leptogenesis\cite{fy}. Within a unified framework, the leptogenesis
constraint favours the nonsupersymmetric model.

The present paper focuses on the general structure of mass matrices and does
not exclude that possible fine tuning could produce a sufficient amount
of baryon asymmetry \cite{ft,branco}, even with nearly diagonal mass matrices.
Other interesting papers on the relation between leptogenesis and mass
matrices are reported in Ref.\cite{other}.

In our simple approach we have not considered the effect of running
masses and mixings from the low scale to the high $M_R$ scale where the
seesaw formula (6) applies.
At our level of approximation, only for
the supersymmetric version with large $\tan \beta$ they can be significant
\cite{ho}. In particular, for spectrum $C_0$ both the double and the single
large mixing are converted to nearly zero mixing,
and for spectra $B_2$ and $C_3$ the double large mixing is converted to
single large mixing. Such spectra are characterized by degeneracy in mass
and sign. However, it has been argued \cite{singh} that the
running of vacuum expectation values could improve the stability of the
lepton mixing matrix.

\end{document}